\documentclass[twocolumn,preprintnumbers,amsmath,amssymb,prl]{revtex4}

\usepackage[dvips]{graphicx}
\usepackage{dcolumn}
\usepackage{bm}
\usepackage[dvips]{epsfig}

\begin{document}

\title{Friction and dilatancy in immersed granular matter.}

\author{T. Divoux and J.-C. G\'eminard.}
\affiliation{Laboratoire de Physique, Ecole Normale Sup\'erieure de
Lyon, CNRS, 46 All\'ee d'Italie, 69364 Lyon cedex 07, France.}

\begin{abstract}

The friction of a sliding plate on a thin immersed granular layer obeys Amonton-Coulomb law. We bring to the fore a large set of experimental results which indicate that, over a few decades of values, the effective dynamical friction-coefficient depends neither on the viscosity of the interstitial fluid nor on the size of beads in the sheared layer, which bears out the analogy with the solid-solid friction in a wide range of experimental parameters. We accurately determine the granular-layer dilatancy, which dependance on the grain size and slider velocity can be qualitatively accounted by considering the rheological behaviour of the whole slurry. However, additional results, obtained after modification of the grain surface by a chemical treatment, demonstrate that the theoretical description of the flow properties of granular matter, even immersed, requires the detailed properties of the grain surface to be taken into account.
\end{abstract}

\maketitle

Conducting studies on immersed granular flows remains of primary interest. A host of geophysical or industrial issues deals with mixtures of grains and fluid as submarine avalanches \cite{Cassar}, snow flows or clay suspensions \cite{coussot}. Also fundamental issues are at stake: one would like to extend the empirical friction law proposed for dense and dry granular flows \cite{GDRmidi} to immersed ones \cite{Pouliquen2}. Enlightening previous studies of sheared and immersed granular media are numerous, and different devices have been developed to describe mixtures of grains and fluids (meaning air or liquids). Among them we choose to focus on the three following.

Studying immersed granular matter flowing down an inclined plane, C. Cassar \textsl{et al.} measured the dynamical friction-coefficient, $\mu$, for different flow configurations \cite{Cassar}. Their results were analyzed using an approach inspired by recent results obtained for dry and dense granular flows \cite{GDRmidi,Pouliquen}: They report the friction coefficient as a function of the dimensionless parameter $I$, first introduced by Da Cruz \textsl{et al.} \cite{DaCruz}, defined to be the ratio of an apt microscopic time scale (inertial, viscous, \ldots) to the relevant macroscopic time scale $\dot \gamma ^{-1}$, where $\dot \gamma$ denotes the shear rate. For immersed granular-matter in the viscous regime \cite{Courrech}, $I \equiv (\dot \gamma \,\eta  _f)/(\alpha \,P_g)$, where $\eta _f$ denotes the viscosity of the intersticial fluid, $P_g$ the pressure exerted on the sheared media and $\alpha$ the normalized permeability of the granular packing \cite{Cassar}. They propose a semi-empirical law for $\mu (I)$ which describes the whole set of data they report for both aerial and immersed granular flows \cite{Cassar,Pouliquen2}. 

Using a Couette geometry Bocquet \textsl{et al.} tuned the pressure within the granular material by applying an upward air flow between the rotating and the stationnary cylinder \cite{Bocquet}. They found out that mean-flow properties and fluctuations in particle motion are coupled. They introduced an hydrodynamic model which quantitatively describes their experiments: The key feature is that the shear force obtained from this model is found to be proportionnal to pressure and approximately independent of the shear velocity. This model does not include any frictional forces between grains, but contains a phenomenological relationship between the viscosity and the dilation of the media.

Using an experimental setup first designed to perform sensitive and fast force-measurements in the dry case \cite{Nasuno}, G\'eminard \textsl{et al.} brought to the fore a dynamic friction-coefficient $\mu$ in the case of an immersed granular layer sheared by means of a sliding plate \cite{Losert}. At low imposed normal-stress, the friction force is shown to be independent of the plate velocity, which holds true as long as the granular material is allowed to dilate \cite{Tardos}. The main difference with the dry case lays in the fact that the slider usually exhibits a continuous sliding instead of the stick-slip motion and in the value of the friction coefficient which is roughly cut down by half \cite{Losert,Geminard2}. The dependance of the frictional coefficient on the fluid viscosity and of the associated dependance of the dilatancy on the slider velocity were not reported. 

Here we report a set of experimental measurements of the friction coefficient and of the dilatancy in a wide range of fluid viscosities and grain sizes
at very low $I$ \cite{Cassar}.
%
Such a study is relevant for several reasons: First, we point out that the quasi-static regime
is unaccessible to the free-surface-flow geometry as size effects crop up in this limit \cite{Pouliquen2}.
In addition, there is a strong discrepancy between the limit of $\mu$ for vanishing $I$ reported in \cite{Cassar}
($\mu\simeq 0,43$ with $I \simeq 4.10^{-3}$)
and those reported for the plane-shear geometry by 
G\'eminard \textsl{et al.} ($\mu\simeq 0.23$, \cite{Losert,Geminard2})
and S. Siavoshi \textsl{et al.} ($\mu\simeq 0.54$, \cite{Kudrolli}), both for $I\simeq 2.10^{-4}$.
How can be explained such discrepancies between those three results?
We also raise the following questions: What does happen when the fluid viscosity or the bead size are changed? How far does the analogy with the Amonton-Coulomb laws remain relevant? What does the effective friction-coefficient depend on? We choose to stick to the canonical plane-shear geometry for which we know that there is a strong analogy between the friction of a slider on an immersed granular layer and the Amonton-Coulomb law \cite{Losert}. In the chosen geometry, the layer is free to dilate which makes it possible to measure both the friction coefficient and the dilation of the granular layer at imposed normal stress. 
\begin{figure}[h]
\begin{center}
\includegraphics[height=5cm ]{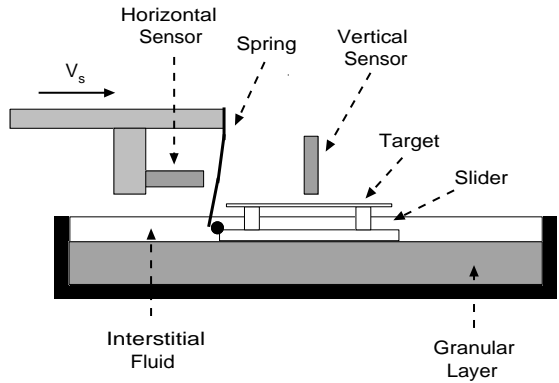}
\end{center}
\caption{\small{\bf Sketch of the experimental setup.}}
\label{fig.schema}
\end{figure}

\textsl{Experimental setup.} - The experimental setup (Fig.~\ref{fig.schema}) is very similar to the one described in \cite{Losert,Geminard2}. A thin plate, the slider, is pushed at the free surface of an immersed granular layer by means of a steel leaf-spring ($k=129 \pm 2$ N.m$^{-1}$) connected to a translation stage driven at constant velocity, $V_s$ by a computer-controlled stepping motor ($V_s$ ranging from 0.1 to 100 $\mu{\rm m.s}^{-1}$). The coupling between the spring and the plate is insured by a metal bead, which avoids applying a torque. The frictional force is monitored by measuring the receding of the spring from its rest position with an inductive sensor (ElectroCorp, EMD1053). The dilatancy is obtained from the vertical displacement of the slider: A second inductive sensor, at rest in the laboratory referential, monitors the distance to a metallic target which endows the slider, which consists in a thin (5 mm) PMMA plate ($76 \times 53$, $53 \times 51$, or $53 \times 24$~mm$^2$). The granular material consists in spherical glass beads (Matrasur Corp.) sieved in order to obtain the three following mean diameters $d = (100 \pm 11)$, $(215 \pm 20)$ and $(451 \pm  40)$ $\mu$m, with a relative standard deviation almost independant of the characteristic grain size. The intersticial fluid consists in distilled water, water and sugar mixtures (viscosity $\eta $ ranging from $1$ to $76$ mPa.s), or Rhodorsil silicon oil (Rhodorsil, viscosity $\eta $ ranging from $71$ mPa.s to $500$ mPa.s). All viscosities were measured, in addition, using an Ubbelohde viscosimeter. The thickness of the granular bed ($6.0$ mm) is always larger than ten bead diameters in order to make sure that the sheared zone is not limited by the bottom of the container and, thus, that edge effects are not at stake \cite{Kudrolli}.
Finally, the contact of the slider with the granular layer is insured by gluing a layer of the largest beads (451 $\mu$m) onto the lower surface. We checked, for a layer of 215 $\mu$m-in-diameter beads, that the friction coefficient is independent of the size of the beads in the glued layer as long as it remains larger than that of the beads in the granular bed (table~\ref{tab.1}, Top).
\begin{table}[h]
 \begin{center}
  \begin{tabular}{cccc}
  \hline
  \hline
  $\o$ ($\mu $m) & 100 & 215 & 451\\
  $\mu _d$ ($\mu $m) & $0.33 \pm  0.02$  &  $0.38 \pm 0.02$ & $0.37 \pm 0.02$ \\
  \hline
  S (mm$^2$) & $53 \times 24$ & $53 \times 51$ & $76 \times 53$ \\
  $\mu _d$ ($\mu $m) & $0.38 \pm  0.02$  &  $0.42 \pm 0.02$ & $0.38 \pm 0.02$ \\
  \hline
  \hline
  \end{tabular}
 \end{center} 
 \caption{\small{Top: Measured friction coefficient $\mu$ as a function of the diameter of the beads that insure the contact at the bottom surface of the slider (The sample consists of $215\, \mu$m beads in water); Bottom: Friction coefficient $\mu$ measured with sliders having different surface area and aspect ratio ($100\, \mu$m beads in a water-sugar mixture, vicosity $\eta = 4.3$ mPa.s.)}}
 \label{tab.1}
\end{table}

\begin{figure}[h]
\begin{center}
\includegraphics[height=6.5cm ]{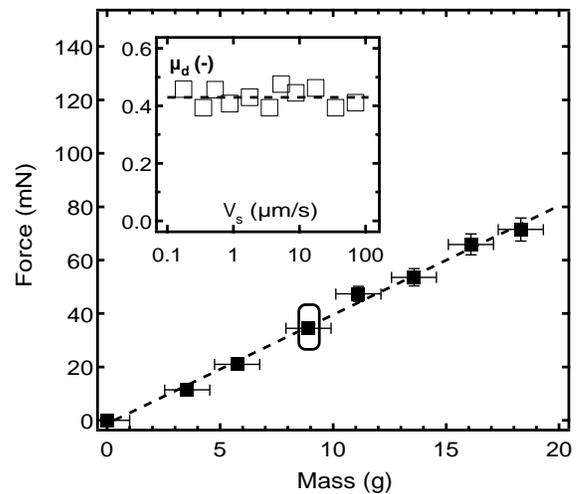}
\end{center}
\caption{\small{{\bf Dynamic frictional force vs. effective mass of the slider}. The effective mass of the slider is obtained by reducing the weight by the buoyancy force. From the slope one can infer $\mu=0.41 \pm 0.02 $ ($V_s = 3.5$ $\mu $m.s$^{-1}$, $451$ $\mu$m beads in silicon oil, viscosity $\eta  = 500$ mPa.s). 
Inset: Friction coefficient vs slider velocity $V_s$ in the same experimental conditions for the mass pointed by the rectangle. The size of the symbols indicates the error bars.}}
\label{fig.FM}
\end{figure}
\textsl{Friction coefficient.} - In our experimental conditions, the value of the spring constant $k$ is chosen so as to observe the continuous motion of the slider in the whole accessible range of the driving velocity $V_s$. After a transient regime, the frictional force reaches a steady-state value which is observed to scale up with the slider mass $m$ (Fig.~\ref{fig.FM}), provided that the buoyancy force is taken into account, and to be independant of the slider velocity $V_s$ (inset, Fig.~\ref{fig.FM}). In addition, we checked that $\mu$ does not significantly depend on the slider surface-area or aspect-ratio (Table~\ref{tab.1}, Bottom), as already known for the dry case \cite{GeLo}. We repeated the procedure for different bead-diameter ($d$ from $100$ $\mu $m to $450$ $\mu $m) and fluid viscosity ($\eta$ from $1$ mPa.s to $500$ mPa.s). We found out that $\mu$ neither depends on $\eta$ nor on $d$ in the whole experimental range. In order to encompass those two results and the independence on the slider velocity $V_s$, we report $\mu$ as a function of the Reynolds number $Re \equiv \rho\,d\,V_s/\eta $, where $\rho$ stands for the fluid density. We estimate $\mu  = 0.38\pm 0.03$ for $10^{-5} \leq I \leq 5.10^{-3}$, which nicely supplement the data reported for the free-surface-flow configuration in \cite{Cassar} that limited to $I\geq 4.10^{-3}$.   
\begin{figure}[h]
\begin{center}
\includegraphics[height=6.5cm]{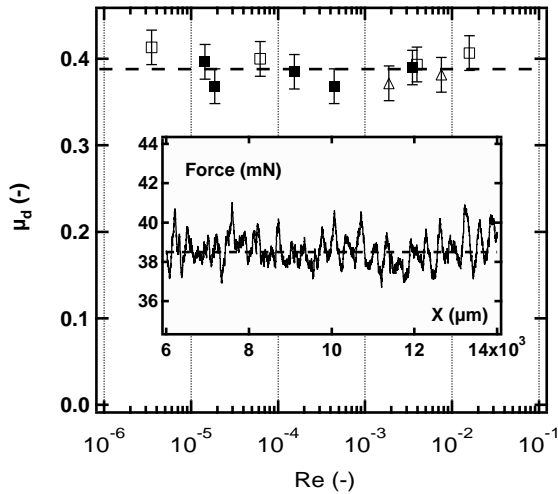}
\end{center}
\caption{
\small{
{\bf Friction coefficient vs. the Reynolds number}. We report data for three different diameters $d$: $\blacksquare$: $100\,\mu $m; $\triangle$: $215\,\mu $m; $\square$: $451\,\mu $m ($\eta $ ranging from $1$ to $500$ mPa.s ). Note that $\mu$ is constant in a range of $Re$ covering more than 4 orders of magnitude. Inset: Frictional force in the steady state regime vs slider position. The dynamic friction-coefficient is defined to be the mean value of the frictional force ($V_s = 8.8 \,\mu$m.s$^{-1}$, $\eta =1$ mPa.s and $m = 10.1$ g).}}
\label{fig.mueff}
\end{figure}

\begin{figure}[h]
\begin{center}
\includegraphics[height=6.5cm]{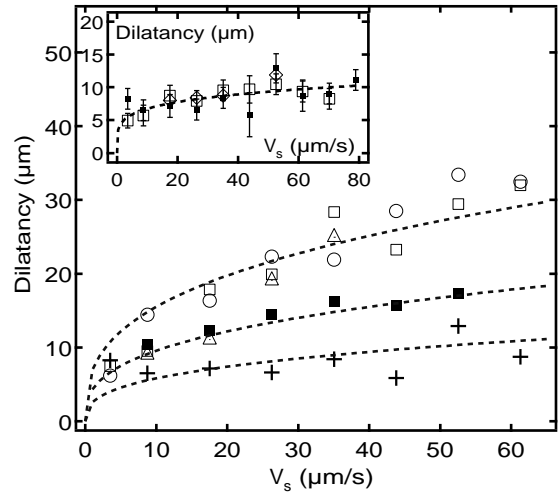}
\end{center}
\caption{
\small{{\bf Total dilation of the layer $\Delta h$ vs. slider velocity $V_s$}.
(symbol, bead diameter, fluid viscosity): ($+$, $100 \,\mu$m, $1$ mPa.s); ($\blacksquare$, $215 \,\mu$m, $1$ mPa.s); ($\triangle$, $451 \,\mu$m, $1$ mPa.s); ($\square $, $451 \,\mu$m, $71$ mPa.s); ($\circ$, $451 \,\mu$m, $500$ mPa.s); The dashed lines correspond to the interpolation of the experimental data to Eq.~\ref{scaling} with $\beta = 2.9\pm 0.3$. Inset: similar results for different reference velocity and normal stress: 
($\blacksquare$, $100 \,\mu$m, $V_{ref}=3.5\,\mu$m.s$^{-1}$); ($\square $, $100 \,\mu$m, $V_{ref}=14 \,\mu$m.s$^{-1}$);
($\diamond $, $100 \,\mu$m, $V_{ref}=3.5 \,\mu$m.s$^{-1}$ and overloaded by $10.0$ g). }}
\label{fig.dilat}
\end{figure}

\textsl{Layer dilation.} - 
As we shall see, in contrast to the frictional coefficient, the dilation of the granular layer is sensitive to the grain size. Experimentally, the moot point consists in obtaining a reproducible reference state. The chosen procedure is as follows:  In order to obtain a well-defined state of compaction, we initially push the slider over a distance of approximately $10$ bead-diameters in the steady regime at a given velocity, henceforth denoted $V_{ref}$ (usually $8.8\, \mu$m.s$^{-1}$, excepted when specified). We then stop the translation stage and move it backwards until the spring goes back to its rest position without loosing contact with the slider, which remains at rest (contact loss could make the slider surf over the granular layer as we push it forth at large velocity, meaning above $40\,\mu $m.s$^{-1}$). We then immediately push the slider forwards at various driving velocities, $V_s$, over a few millimeters and monitor the vertical position of the plate. We observe that the total variation of the vertical position of the plate $\Delta h$, or total dilation, increases with the bead diameter $d$ and the velocity $V_s$. By constrast, $\Delta h$ does not significantly depend on the interstitial-fluid viscosity $\eta$  (Fig.~\ref{fig.dilat}). We checked that these latter measurements neither depend on the preparation of the granular layer (by varying the velocity of reference $V_{ref}$), nor on the slider mass $m$ (Fig.~\ref{fig.dilat}, inset).

\textsl{Discussion.} - The dependence of the total dilation on the velocity and on the bead diameter can be accounted by the two following ingredients: First, we can guess that, due to the steric interaction between the grains (solid contact between grains or hydrodynamical interaction), the local shear-stress $\sigma_s$ induces a local normal-stress $\sigma_n = \alpha(\phi)\,\sigma_s$. In a first approximation, the coefficient $\alpha$, which describes a geometrical property, depends only on the volume fraction of the grains, $\phi$, and not on the shear rate, $\dot\gamma$. The assumption is correct, at least for a dense suspension in the limit of small $\dot\gamma$ \cite{huang05}. We point out that, from this local relation between $\sigma_n$ and $\sigma_s$, we recover the apparent friction law, $F = \mu \,m g$ where $\mu = 1/\alpha$, independant on the shear rate $\dot\gamma$ or on the surface area $S$ of the slider provided that $\alpha$ does not significantly depend on $\phi$ \cite{huang05}. Second, we assume, as already proposed by Bocquet {\it et al} \cite{Bocquet}, that the rheological behavior of the immersed granular-material can be accounted, in addition, by $\sigma_s = \eta(\phi) \,\dot\gamma$, where the effective viscosity $\eta$ diverges algebraically as a function of $\phi$ near a critical volume fraction $\phi_c$: $\eta  = {\eta _0}/{\left(1-\phi/\phi _c\right)^{\beta}}$. At this point, assuming that the total dilation $\Delta h$ is mainly due to the dilation of a constant number $N$ of layers underneath the slider and linearizing the velocity profile in this region where the dilation is the larger, we write $\dot\gamma = V_s/(Nd)$ and get the following scaling law: 
\begin{equation}
\Delta h \propto {V_s}^{1/\beta}{d^{(\beta-1)/\beta}}
\label{scaling}
\end{equation} 
The interpolation of the experimental data reported in Fig.~\ref{fig.dilat} leads to $\beta = 2.9 \pm 0.3$, which compares quantitatively with the values reported in \cite{Bocquet} and references therein. 

Thus, our experimental measurements are in agreement with the conclusions of references \cite{Cassar} and \cite{Bocquet}. However, the discrepancy with the values issued in \cite{Losert,Geminard2} and \cite{Kudrolli} remains unexplained and deserves to be discussed.
  Few physical origins can be taken responsible for such a dispersion related in the literature. Among them, in the case of spherical glass-beads, the polydispersity of the batches and the properties of the beads surface, especially its roughness. On the one hand, using unsieved samples (polydispersity about 25\%), we measured significantly higher values $\mu \simeq 0.47\pm 0.02$ of the friction coefficient, which provides us with a rough estimate of the polydispersity effects. For all the data reported above, we use sieved samples, the standard deviation in the diameter being about $10\%$ as in references \cite{Losert,Geminard2,Kudrolli}, so that we estimate that the polydispersity is not enough for explaining the observed discrepancy. On the other hand, only a few observations, dealing with the influence of the surface properties on the effective friction coefficient, have been reported: It has been recently proven that the roughness can drastically alter the dynamic angles of repose for dry materials \cite{Ottino} and even be a motor for shear-induced segregation in immersed granular-materials \cite{Pouligny}.
  The beads used in our study and in \cite{Kudrolli,Losert,Geminard2} may not present the same surface properties. In order to test the influence of the bead surface, we altered $451 \,\mu$m diameter beads by immersing them for $30$ min in a $1.0$ mol.L$^{-1}$ sodium-carbonate solution \cite{Herve}. The beads were then thoroughly washed up using distilled water and $\mu$ immediatly assessed. We obtained $\mu = 0.30 \pm 0.01$ and, thus,  a significant decrease of about 20\% by changing only the surface properties, even in immersed granular matter. We are currently performing an extensive and careful study of the friction coefficient as a function of surface roughness of the beads. 
  
As a conclusion, our measurements of the friction coefficient nicely supplement the results reported by Cassar {\it et al} \cite{Cassar} and we interpret our measurements of the dilatancy in the framework of the hydrodynamical model proposed by Bocquet \textsl{et al.} \cite{Bocquet}. The friction coefficient $\mu$ remains constant over a large range of $Re$ values because the dilatancy of the layer is a free parameter that adapts as the slurry is sheared at different velocities.
In this regime, the effective friction-coefficient depends neither on the fluid viscosity nor on the bead diameter. However, we underscore that the properties of the grain surface play an important role in the rheological properties of immersed granular-matter and could be responsible for the dispersion of $\mu $ values encountered in the literature. Finally, we emphasize that measurements of $\mu $, defined from the mean value of the friction force in the steady-state regime, does not provide any piece of information about the grain- or fluid-characteristics. 
As an extension of this work, we are currently focusing on the fluctuations of the frictional force in the steady-state regime (inset, Fig.~\ref{fig.mueff}), from which we hope to recover a signature of the components of the slurry.
 

\end{document}